\documentclass[slac_one]{revtex4}
\usepackage{graphicx}
\usepackage{fancyhdr}
\usepackage{subfigure} % subfigures in 1 figure
\graphicspath{{./}}
\def\integLumi{315~nb$^{-1}$}
\def\integLumiUncertainty{$\pm 11$\%}
\def\MjjLowerExclusionTeV{0.30} % Note: no units attached!
\def\MjjUpperFitRange{1900} % Note: no units attached!
\def\MjjLowerLimitMRSTTeV{1.26} % Note: no units attached!
\def\MjjLowerLimitMRSTEXPECTEDTeV{1.06} % Note: no units attached!
\def\pTcut{80} % Note: no units attached!
\def\GeVcc{${\rm GeV}$}
\def\TeVcc{${\rm TeV}$}
\def\GeVc{${\rm GeV}$}

\begin{document}

%Title of paper
\title{Search for New Particles in 2-Jet Final States in 7~TeV 
  Proton-Proton Collisions with the ATLAS Detector at the LHC} %% Paper title goes here

% Repeat the \author .. \affiliation  etc. as needed
%
% \affiliation command applies to all authors since the last
% \affiliation command. The \affiliation command should follow the
% other information

\author{S.~L.~Cheung, on behalf of the ATLAS Collaboration}
\affiliation{The University of Toronto, Toronto, ON M5S 1A7, Canada}

\begin{abstract}
  A search for new heavy particles manifested as resonances in two-jet
  final states is presented.  The data were produced in $\sqrt{s}=7~TeV$ 
  proton-proton collisions by the Large Hadron Collider (LHC) and
  correspond to an integrated luminosity of \integLumi~collected by
  the ATLAS detector.  No resonances were observed.  Upper limits were
  set on the product of cross section and signal acceptance for
  excited-quark ($q^*$) production as a function of $q^*$ mass.  These
  exclude at the 95\%~CL the $q^*$ mass interval
  $\MjjLowerExclusionTeV < m_{q^*} < \MjjLowerLimitMRSTTeV$~\TeVcc,
  extending the reach of previous experiments.
\end{abstract}

%\maketitle must follow title, authors, abstract
\maketitle

\thispagestyle{fancy}

% body of paper here - Use proper section commands
% References should be done using the \cite, \ref, and \label commands
% Put \label in argument of \section for cross-referencing
%\section{\label{}}

%%%-----------------------------------------------------------------------
\section{Introduction}
Two-jet (dijet) events in high-energy proton-proton ($pp$) collisions
are usually described in the Standard Model (SM) by applying quantum
chromodynamics (QCD) to the scattering of beam-constituent quarks and
gluons.  Several extensions beyond the SM predict new heavy particles,
accessible at LHC energies, that decay into two energetic
partons. Such new states may include an excited composite quark $q^*$,
exemplifying quark
substructure~\cite{refTherein}; an
axigluon predicted by chiral color
models~\cite{refTherein}; a flavour-universal
color-octet coloron~\cite{refTherein}; or a
color-octet techni-$\rho$ meson predicted by models of extended
technicolor and topcolor-assisted
technicolor~\cite{refTherein}.

The dijet invariant mass ($m^{jj}$) observable is particularly
sensitive to such new objects.  At the Fermilab Tevatron collider,
1.13~fb$^{-1}$ of $p\bar{p}$ collision data were used to exclude the
existence of excited quarks $q^*$ with mass $260 < m_{q^*} <
870$~\GeVcc~\cite{refTherein}.  This analysis focused on a search for
the excited quarks because of the accessible predicted cross
section~\cite{refTherein}\ for such particles and the benchmark nature
of the model that allows limits on acceptance times cross section to
be set for resonant states with intrinsic widths narrower than the
experimental resolution.

%%%-----------------------------------------------------------------------
\section{Measurement and Exotic Search}
\subsection{Jet Reconstruction}
Jets~\cite{atlasjet} were reconstructed in ATLAS~\cite{atlas} using
the anti-$k_T$ jet clustering algorithm~\cite{refTherein} with a
size parameter of $R=0.6$.  The inputs to this algorithm were 3D
topological clusters of calorimeter cells seeded by cells with energy, 
calibrated at electromagnetic scale, significantly above the measured
noise. Jet 
four-vectors were constructed by performing a four-vector sum over
these cell clusters and calibrated by MC-derived $p_T$-dependent and
$\eta$-dependent calibration factors.   The dijet mass observable
$m^{jj}$ was computed without unfolding jets to hadrons or partons.

%%%-----------------------------------------------------------------------
\subsection{Event Selection}
\begin{figure}%[h]
  \begin{center}
    \includegraphics[width=60mm]{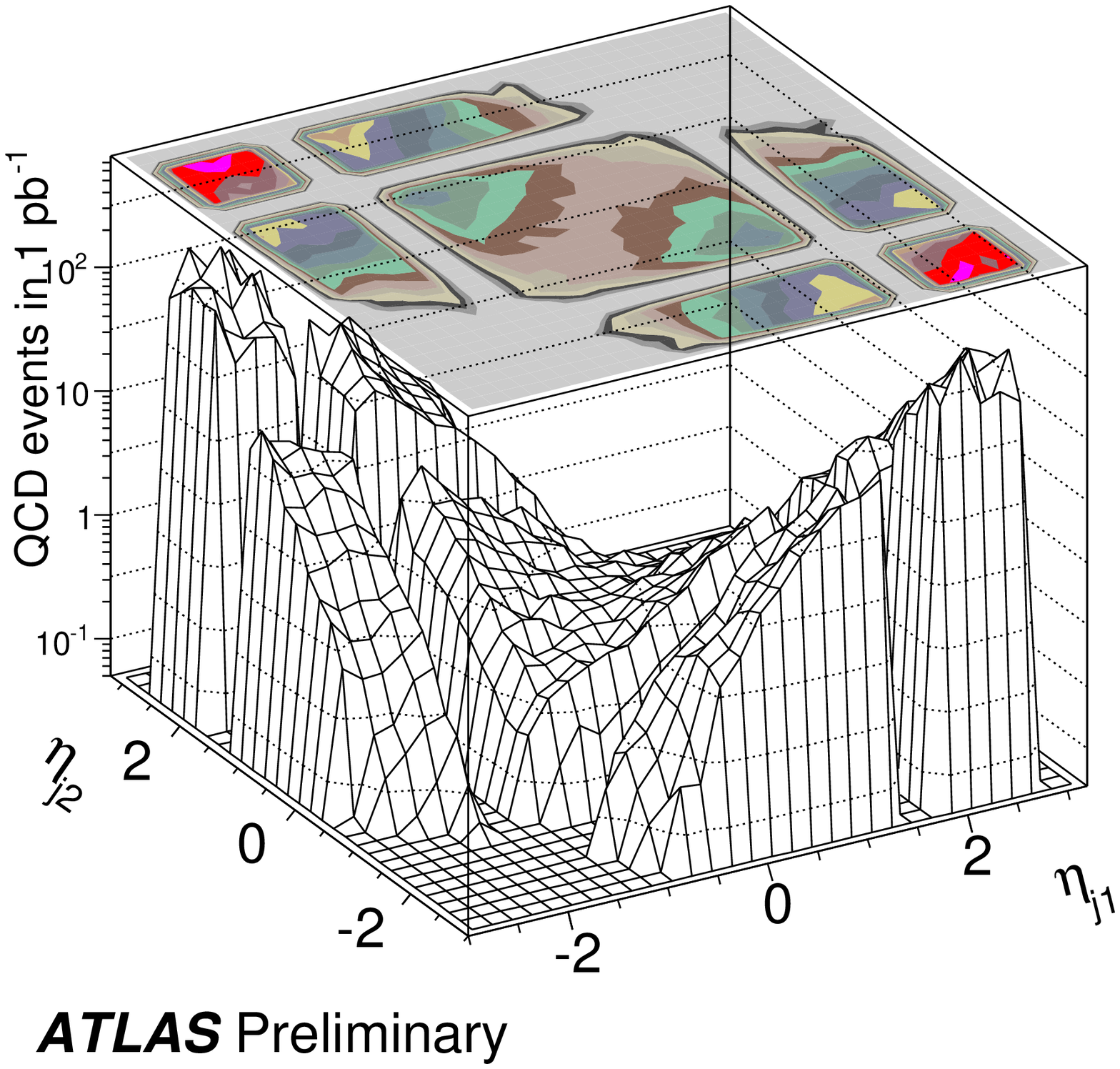}%placeHolder.eps}
    \includegraphics[width=60mm]{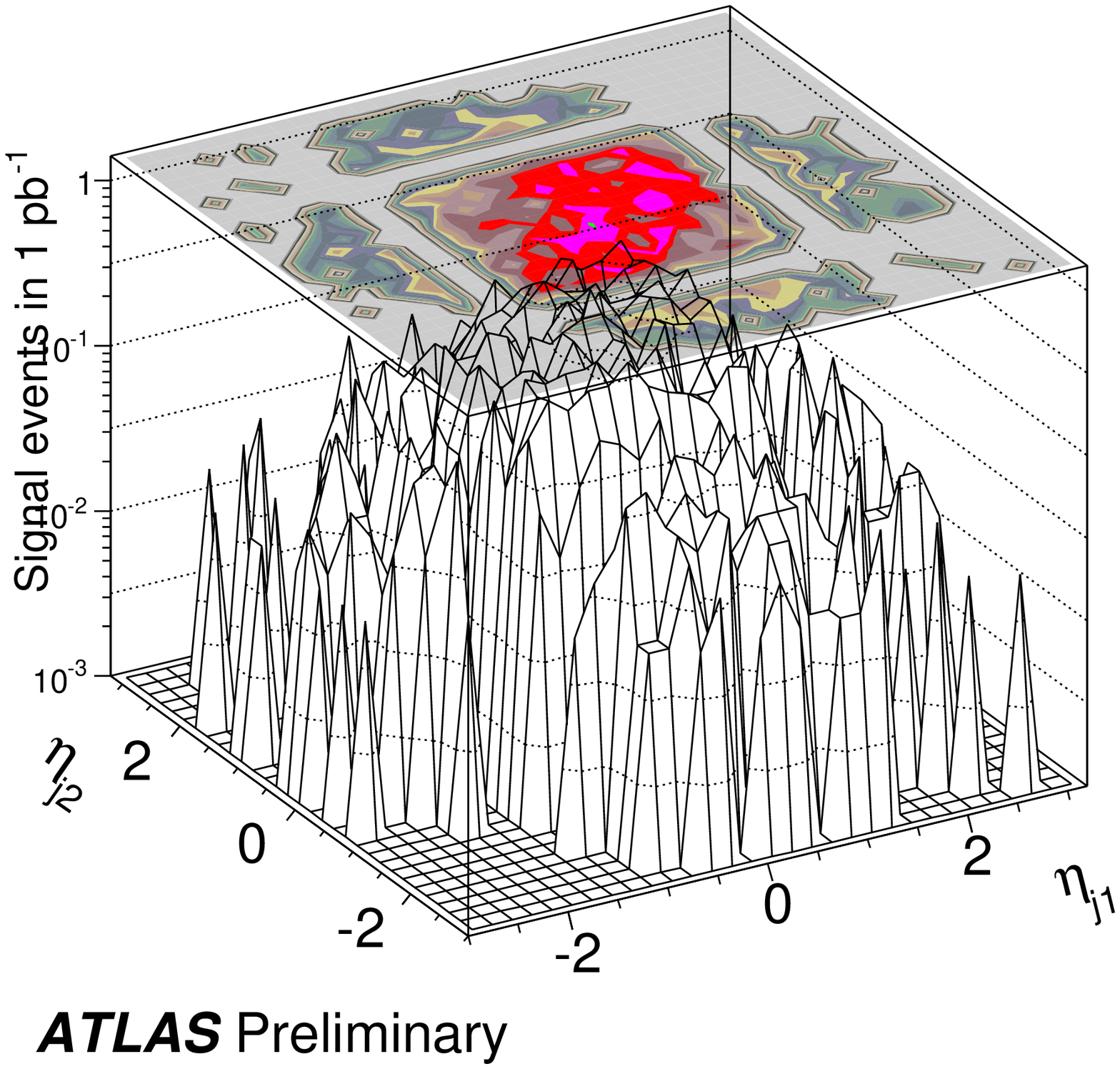}%placeHolder.eps}
  \end{center}
  \caption{Surface plots describing the expected distributions of event
    yields in the observables $\eta_{j_1}$ and $\eta_{j_2}$ for dijets
    with $875 \leq m^{jj} \leq 1020$~GeV, a range defined by the
    predicted $\pm 1\sigma$ region of a 1~TeV excited quark $q^*$,
    identified in MC [left] QCD background and [middle] 1~TeV $q^*$
    signal samples.}
  \label{fig:optimizeEta}
\end{figure}

Events were required to contain at least one primary collision vertex
(defined by at least five reconstructed charged-particle tracks) and
at least two jets after the criteria that the leading jet $p_T^{j_1}$
and the subleading jet $p_T^{j_2}$ are larger than~$\pTcut$~\GeVc\ and
$30$~\GeVc\ respectively.
The events with a poorly measured jet~\cite{jetdq} of $p_T >
15$~\GeVc\ were vetoed to prevent from swapping between the
poorly-measured jet and one of the 2 leading jets.  The two leading
jets were required to satisfy several quality
criteria~\cite{refTherein} and to lie outside the crack region, i.e.,
$1.3 < |\eta^{\rm jet} | < 1.8$.  Finally, both jets were required to
be within $\left|\eta^{\rm jet}\right| < 2.5$ as well as
$\left|\eta^{j_1}-\eta^{j_2}\right| < 1.3$ in order to maximize the
signal discovery potential illustrated in Fig.~\ref{fig:optimizeEta}.

%%%-----------------------------------------------------------------------
\subsection{Background Determination and Statistical Test}
The background shape was determined by fitting the observed spectrum
with the function~\cite{refTherein}
\begin{equation}
f(x) = p_1 (1 - x)^{p_2} x^{p_3 + p_4\ln x},
\label{eq:f}
\end{equation}
where $x\equiv m^{jj}/\sqrt{s}$  and $p_{\{1,2,3,4\}}$ are free
parameters.  This smooth and monotonic form of Eqn.~\ref{eq:f}
demonstrated a good agreement with QCD-predicted dijet mass
distributions in $pp$\ collisions at $\sqrt{s} = 7$~TeV, as evidenced
by a $\chi^2 = 27$ for 22 degrees of freedom over the dijet mass range
$200 < m^{jj} < \MjjUpperFitRange$~\GeVcc.  This supported the use of
Eqn.~\ref{eq:f} to estimate the background shape in the observed
$m^{jj}$ distribution.  The results of fitting the data with
Eqn.~\ref{eq:f}\ are shown in Fig.~\ref{fig:spectrumAndFit}.

Six statistical tests of the background-only hypothesis were employed to
determine the presence or absence of detectable $m^{jj}$ resonances in
this distribution: 
the BumpHunter~\cite{refTherein}, 
the Jeffreys divergence~\cite{refTherein}, 
the Kolmogorov-Smirnov test, 
the likelihood,
the Pearson $\chi^2$, 
and the TailHunter statistic~\cite{refTherein}. All these tests
were consistent with the conclusion that the fitted parameterization
described the observed data distribution well, with $p$-values in
excess of 51\%.  These observations supported the background-only
hypothesis.  In the absence of any observed discrepancy with the
zero-signal hypothesis, a Bayesian approach was used to set 95\%
credibility-level (CL) upper limits on $\sigma\cdot {\cal A}$ for
hypothetical new particles decaying into dijets with $\left|\eta^{\rm
  jet}\right| < 2.5$.

%%%-----------------------------------------------------------------------
\subsection{Systematic Uncertainty and Limit Setting}
The dominant sources of systematic uncertainty were associated with
the absolute jet energy scale (JES), the background
fit parameters and  the integrated luminosity.  The
JES uncertainty was quantified as a function of $p_T$ and $\eta^{\rm
  jet}$, with values in the range $6\sim9\%$~\cite{atlasJES}.  The
systematic uncertainty on the background determination was taken from
the uncertainty on the fit parameters of Eqn.~\ref{eq:f} to the data
sample.  The uncertainty on $\sigma\cdot {\cal A}$ due to integrated
luminosity was estimated to be
\integLumiUncertainty~\cite{refTherein}.  All these effects were
incorporated as nuisance parameters into the likelihood function 
and then marginalized by numerically integrating the product of this
modified likelihood, the prior in signal cross section, and the priors
corresponding to the nuisance parameters to arrive at a modified posterior
probability distribution.

\begin{figure}
  \begin{center}
    \subfigure[]{
      \includegraphics[width=60mm]{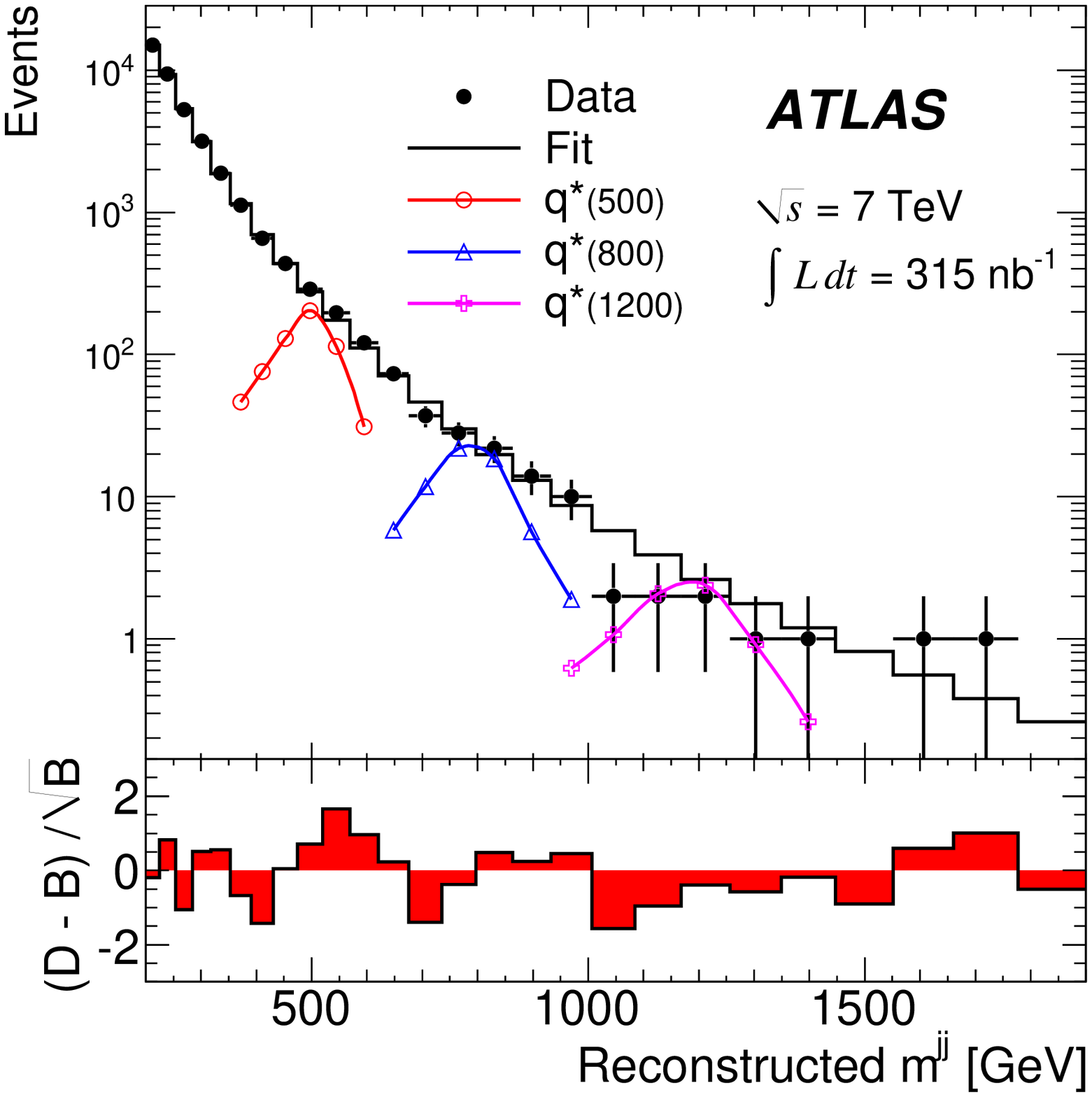}
      \label{fig:spectrumAndFit}
    }
    \subfigure[]{
      \includegraphics[width=60mm]{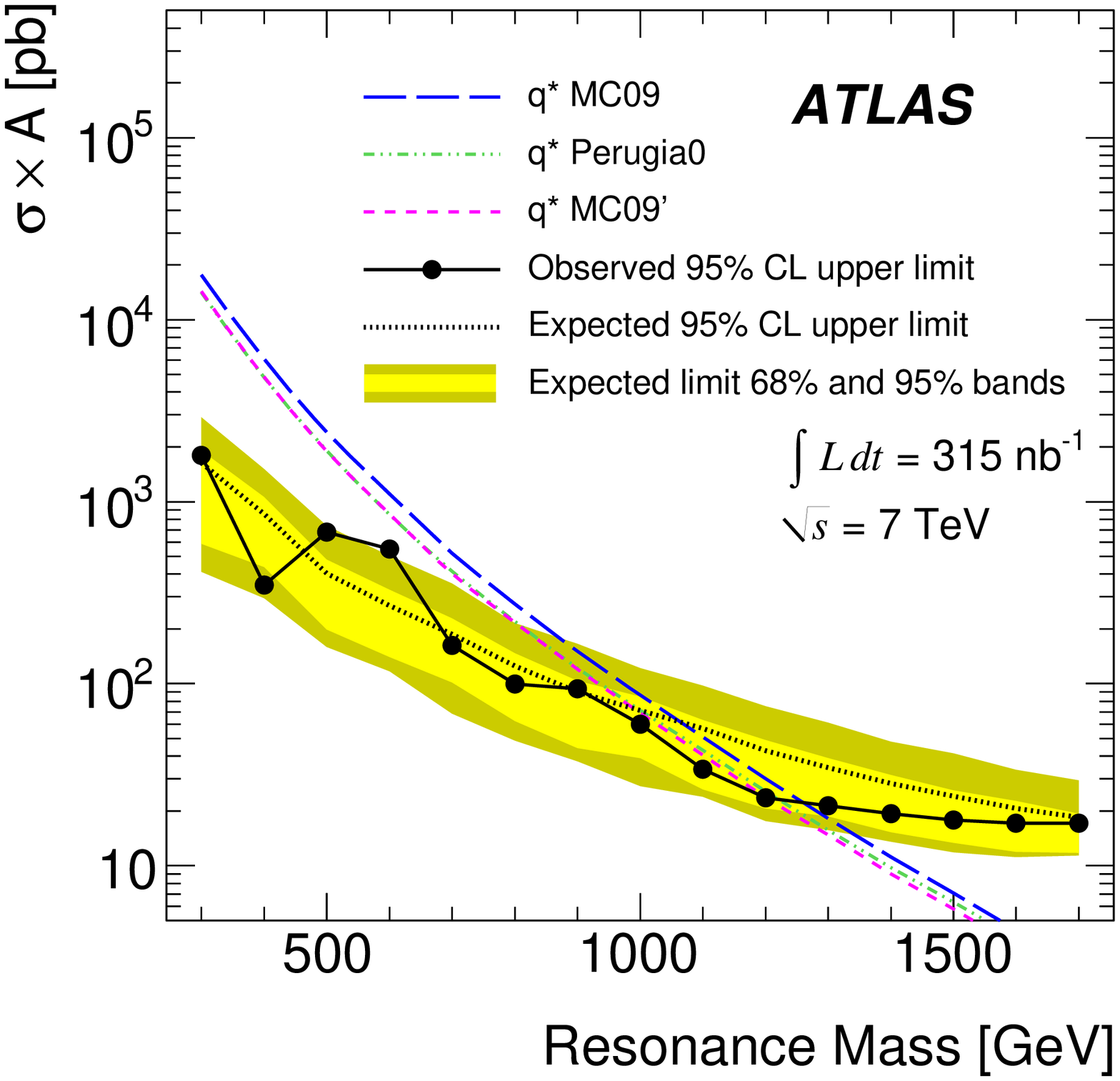}
      \label{fig:ExclusionWithStatSystematics}
    }
  \end{center}
  \caption{[Left]  The data (D) dijet mass distribution (filled points)
    fitted using a binned background (B) distribution described by
    Eqn.~\ref{eq:f} (histogram).  The predicted $q^*$
    signals~\cite{refTherein} for excited-quark masses of
    500, 800, and 1200~\GeVcc\ are overlaid, and the bin-by-bin
    significance of the data-background difference is shown. 
    [Right]  The 95\% CL upper limit on $\sigma\cdot {\cal A}$ as a
    function of dijet resonance mass (black filled circles).  The
    black dotted curve shows the expected 95\% CL upper limit and the
    light and dark yellow shaded bands represent the 68\% and 95\%
    credibility intervals of the expected limit, respectively.  The
    dashed curves represent excited-quark $\sigma\cdot {\cal A}$
    predictions for different MC tunes, each using a different PDF
    set. }
\end{figure}

Fig.~\ref{fig:ExclusionWithStatSystematics} depicts the resulting
95\% CL upper limits on $\sigma\cdot {\cal A}$ as a function of the
$q^*$ resonance mass after the incorporation of systematic
uncertainties. The corresponding observed 95\% CL excited-quark mass
exclusion region was found to be $\MjjLowerExclusionTeV < m_{q^*} <
\MjjLowerLimitMRSTTeV$~\TeVcc\/~\cite{firstPRL}, with the expected
exclusion range of $\MjjLowerExclusionTeV < m_{q^*} <
\MjjLowerLimitMRSTEXPECTEDTeV$~\TeVcc, using MRST2007 PDFs in the 
ATLAS default MC09 tune.

%%%-----------------------------------------------------------------------
\section{Conclusion}
A model-independent search for new heavy particles manifested as mass
resonances in dijet final states was conducted using a
\integLumi~sample of 7~TeV proton-proton collisions recorded by the
ATLAS detector.  No evidence of a resonance structure was found and
upper limits at the 95\% CL were set on the products of cross section
and signal acceptance for  hypothetical new $q^*$ particles decaying
to dijets.  These data exclude at the 95\% CL excited-quark masses
from the lower edge of the search region,
\MjjLowerExclusionTeV~\TeVcc, to \MjjLowerLimitMRSTTeV~\TeVcc\ for a
standard set of model parameters and using the ATLAS default MC09
tune~\cite{refTherein}.  In the future, such searches
will be extended to exclude or discover additional hypothetical
particles over greater mass ranges.

%%%-----------------------------------------------------------------------

\end{document}